\documentclass[preprint,aps,prd,showpacs,groupedaddress,superscriptaddress,%
floatfix,nofootinbib]{revtex4}%
\usepackage{graphicx}
\usepackage{amsmath}
\usepackage{bm}
\begin{document}
\title{Cornell Potential Parameters for $\bm{S}$-wave Heavy Quarkonia 
}
\author{Hee~Sok~Chung}
\affiliation{Department of Physics, Korea University, Seoul 136-701, Korea}
\author{Jungil~Lee}
\affiliation{Department of Physics, Korea University, Seoul 136-701, Korea}
\author{Daekyoung~Kang}
\affiliation{Physics Department, Ohio State University, 
Columbus, Ohio 43210, USA}

\begin{abstract}
We compute derived quantities for various values of the model parameter
of the Cornell potential
model for the $S$-wave heavy quarkonia with radial quantum numbers $n=1$,
$2$, and $3$. Our results can be used to determine leading
and relative-order-$v^2$ nonrelativistic quantum chromodynamics matrix
elements for $S$-wave charmonia and bottomonia such as $\psi(2S)$,
$\eta_c(2S)$, and $\Upsilon(nS)$ for $n=1$, $2$, and $3$. These matrix
elements will be essential ingredients for resumming relativistic
corrections to processes involving those $S$-wave heavy quarkonium states.
\end{abstract}

\maketitle

\section{Introduction\label{intro}}

One of the most interesting recent developments in heavy-quarkonium 
phenomenology is the introduction of a new technique for resumming
relativistic corrections to $S$-wave quarkonium production
and decay rates. The technique resums corrections to all orders in the 
heavy-quark velocity $v$ in the heavy-quark-antiquark ($Q\bar{Q}$)
rest frame~\cite{Bodwin:2006dn,Bodwin:2007fz} within the color-singlet
mechanism of the nonrelativistic quantum chromodynamics (NRQCD)
factorization approach~\cite{Bodwin:1994jh}. With the method, 
the relative-order-$v^2$ NRQCD matrix element for the $S$-wave
charmonium, which has a power-ultraviolet divergence and needs
subtraction, has been evaluated with improved
accuracy~\cite{Bodwin:2006dn,Bodwin:2007fz}
compared with lattice calculations~\cite{bks} and with the 
determination~\cite{Braaten:2002fi}
obtained by using the Gremm-Kapustin relation~\cite{Gremm:1997dq}.
In Ref.~\cite{Bodwin:2006dn}, the generalized version of the 
Gremm-Kapustin relation within the Cornell 
potential model~\cite{Eichten:1978tg} was derived.
The generalized Gremm-Kapustin relation allows one to resum a 
class of relativistic corrections in a potential-model color-singlet
$Q\bar{Q}$ wave function. The resummation method has also been applied 
to determine leading-order NRQCD matrix elements for the $1S$ charmonium 
states. The resultant values for the matrix elements are significantly
greater than those known previously~\cite{Bodwin:2006dn,Bodwin:2007fz}. 
In addition, the method has provided a reasonable 
solution~\cite{Bodwin:2007ga} to the long-standing puzzle 
of the cross section for $e^+e^-\to J/\psi+\eta_c$
measured at the $B$-factories~\cite{puzzle},
which has been one of the greatest discrepancies between theory and
experiment within the Standard Model.

Therefore, it is worthwhile to extend this method to study radially
excited $S$-wave charmonia, $\psi(2S)$ and $\eta_c(2S)$, and 
spin-triplet $S$-wave bottomonia, $\Upsilon(nS)$, with radial
quantum numbers $n=1$, $2$, and $3$. Unfortunately, the potential-model
parameters and derived quantities reported in 
Ref.~\cite{Bodwin:2006dn} are only for the $1S$ and the $2S$ states.
The tabulation of the parameters in Ref.~\cite{Bodwin:2006dn} is
not convenient to use in combination with the improved version of
the resummation method reported in Ref.~\cite{Bodwin:2007fz}, which
considered only the $1S$ charmonium states.

In this paper, we compute derived quantities for various values of the 
model parameter
of the Cornell potential model for $S$-wave heavy quarkonia 
with radial quantum numbers $n=1$, $2$, and $3$.  The Cornell potential
is a linear combination of the Coulomb and linear potentials. 
Following Ref.~\cite{Eichten:1978tg}, we rescale the Schr\"odinger
equation and solve the equation numerically
for various values of the model parameter $\lambda$, which determines
the strength of the Coulomb potential relative to the linear potential.
To achieve improved accuracies in calculating energy eigenvalues, we use
the numerical method\footnote{
In previous calculations in Refs.~\cite{Bodwin:2006dm,Bodwin:2006dn},
the authors quoted the energy eigenvalues determined in 
Ref.~\cite{Eichten:1978tg}. A recently improved analysis given in
Ref.~\cite{Bodwin:2007fz} employed the numerical method used in the 
present work.}
given in Refs.~\cite{Kang:2006jd,Aichinger:2005}.
The potential-model parameters and derived quantities are listed as
functions of $\lambda$. These parameters are ready for use to
determine leading and relative-order-$v^2$ color-singlet NRQCD matrix
elements for both $S$-wave charmonia and bottomonia. 

The remainder of this paper is organized as follows.
In Sec.~\ref{sec:model}, we give a brief description of the Cornell 
potential model. Our numerical results are given in Sec.~\ref{sec:result},
which is followed by a discussion in Sec.~\ref{sec:discussion}.

\section{Potential Model\label{sec:model}}
In order to compute leading and higher-order color-singlet NRQCD 
matrix elements for an $S$-wave heavy quarkonium with the radial
quantum number $n=1$, $2$, or $3$, we need to compute the binding 
energy $\epsilon_{nS}$ of the $nS$ state that appears in the 
generalized Gremm-Kapustin relation~\cite{Bodwin:2006dn,Gremm:1997dq}. 
In this section, we describe briefly the Cornell potential 
model~\cite{Eichten:1978tg} that we use to compute $\epsilon_{nS}$.
We refer the reader to Refs.~\cite{Bodwin:2006dn,Bodwin:2007fz} 
which contain more complete descriptions of the model in conjunction
with the NRQCD factorization formalism and the resummation technique
for relativistic corrections to quarkonium processes.

We employ the Cornell potential~\cite{Eichten:1978tg}, 
which parametrizes the $Q\bar Q$ potential $V(r)$ as a linear combination
of the Coulomb and linear potentials:
\begin{equation}
V(r)=-\frac{\kappa}{r}+\sigma r,
\label{model-V}%
\end{equation}
where $\kappa$ is a model parameter for the Coulomb 
strength and $\sigma$ is the string tension. The relation between
the string tension $\sigma$ and the corresponding parameter $a$
in the original formulation of the Cornell potential model 
\cite{Eichten:1978tg} is $a=1/\sqrt{\sigma}$.
By varying the parameters in the Cornell potential, one can obtain good 
fits to lattice measurements of the $Q\bar Q$ static potential 
\cite{Bali:2000gf}. Therefore, we assume that the use of the Cornell 
parametrization of the $Q\bar Q$ potential should result in errors that are 
much less than the order-$v^2$ errors (about 30\% for a charmonium and
about 10\% for a bottomonium) that are inherent in the leading-potential
approximation to NRQCD~\cite{Bodwin:2006dn,Bodwin:2007fz}.

The Schr\"{o}dinger equation for the $S$-wave radial wave function 
$R_{nS}(r)$ with the radial quantum number $n$ is 
\begin{equation}
\left[
-\frac{1}{mr^2} \frac{d}{dr}\left( r^2 \frac{d}{dr}\right)
+V(r)
\right]R_{nS}(r)=\epsilon_{nS} R_{nS}(r),
\label{radial}%
\end{equation}
where $m$ is the quark mass and $\epsilon_{nS}$ is the binding energy for 
the $nS$ state. The model parameter $m$ is distinguished
from the heavy-quark mass $m_Q$ that appears in the short-distance
coefficients of NRQCD factorization formulas. For an $S$-wave state,
the wave function is $\psi_{nS}(r)=R_{nS}(r)/\sqrt{4\pi}$.

The Schr\"{o}dinger equation in Eq.~(\ref{radial}) depends on 
the model parameters $m$ and $\kappa$, where
we assume that the string tension $\sigma$ is common to both 
charmonium and bottomonium states and that the value for $\sigma$ can be 
determined from lattice measurements of the $Q\bar{Q}$ static
potential.
Then, the dependence on the flavor appears through $m$ and $\kappa$.
Introducing the scaled radius $\rho$ and scaled coupling
$\lambda$~\cite{Eichten:1978tg},
\begin{subequations}
\begin{eqnarray}
\rho&=&(\sigma m)^{1/3}\, r,
\label{r-rho}%
\\
\lambda&=&\frac{\kappa}{(\sigma/m^2)^{1/3}},
\label{kappa-lambda}%
\end{eqnarray}
\label{dimensionless}%
\end{subequations}
which are dimensionless, one can rewrite the radial equation 
in Eq.~(\ref{radial}) as~\cite{Eichten:1978tg}
\begin{equation}
\left[
\frac{d^2}{d\rho^2}
+\frac{\lambda}{\rho}-\rho+\zeta_{nS}
\right]u_{nS}(\rho)=0,
\label{eq:u}%
\end{equation}
where $u_{nS}(\rho)$ and $\zeta_{nS}$ are the dimensionless radial
wave function and the dimensionless energy eigenvalue of the $nS$
state, respectively. The relation between $R_{nS}(r)$ and 
$u_{nS}(\rho)$ is
\begin{equation}
\label{eq:psir}%
R_{nS}(r)=
\sqrt{\sigma m}
\, \frac{u_{nS}(\rho)}{\rho},
\end{equation}
where the wave functions are normalized according to
\begin{equation}
\int_0^\infty |u_{nS}(\rho)|^2 d\rho
=\int_0^\infty |R_{nS}(r)|^2 r^2dr=1.
\end{equation}
The binding energy is related to the dimensionless eigenvalue 
$\zeta_{nS}$ as
\begin{equation}
\epsilon_{nS}= 
[\sigma^2/m]^{1/3}\zeta_{nS}(\lambda).
\label{eq:ezeta}%
\end{equation}

Note that Eq.~(\ref{eq:u}) depends only on $\lambda$. Therefore,
the scaled equation can be solved for a given $\lambda$ to get
the wave function $u_{nS}(\rho)$ and the eigenvalue $\zeta_{nS}$.
The flavor dependence appears when we invert them to get the radial
wave function $R_{nS}(r)$ and the energy eigenvalue $\epsilon_{nS}$.
In this step, $m$ looks independent of the model parameter $\lambda$.
However, we can express $m$ in terms of $\sigma$, $\lambda$, 
and the $1S$-$2S$ mass 
splitting~\cite{Eichten:1978tg,Bodwin:2006dn,Bodwin:2007fz}:
\begin{equation}
m (\lambda)=
\sigma^2 \left[\frac{\zeta_{2S}(\lambda)-\zeta_{1S}(\lambda)}
                    {m_{2S}-m_{1S}}\right]^3.
\label{m-lam}%
\end{equation}
For $S$-wave states, the wave function at the origin
$\psi_{nS}(0)=R_{nS}(0)/\sqrt{4\pi}$ can be expressed
as~\cite{Eichten:1978tg,Bodwin:2006dn,Bodwin:2007fz}
\begin{equation}
|\psi_{nS}(0)|^2
=\frac{m}{4\pi}\int d^3r
   |\psi_{nS}(r)|^2
   \frac{\partial V(r)}{\partial\, r} 
=\frac{\sigma\, m(\lambda)}{4\pi}
\left[
1+\lambda F_{nS}(\lambda)
\right]
,
\label{eq:psi0rhom2}%
\end{equation}
where $F_{nS}(\lambda)$ is the expectation value of $1/\rho^2$ 
for the $nS$ state:
\begin{equation}
F_{nS}(\lambda)=
\int_0^\infty \frac{d\rho}{\rho^2} \,\left|u_{nS}(\rho)\right|^2.
\label{eq:rhom2}%
\end{equation}

For purposes of computation of the NRQCD matrix elements, it is
convenient to express those matrix elements in terms of the
potential-model parameters listed above.
A convenient parametrization can
be found in Ref.~\cite{Bodwin:2007fz}. 
According to Ref.~\cite{Bodwin:2007fz}, 
the leading-order NRQCD matrix element for the $S$-wave 
heavy quarkonium depends on $\lambda$, $m(\lambda)$, 
and $F_{nS}(\lambda)$.
The ratio of the relative-order-$v^2$ NRQCD matrix element to the 
leading-order one is proportional to $\zeta_{nS}(\lambda)$.
The electromagnetic decay rate for the quarkonium, in which 
relativistic corrections to all orders in $v$ are resummed, is 
expressed in terms of these NRQCD matrix elements~\cite{Bodwin:2007fz}.
Because $m(\lambda)$, $F_{nS}(\lambda)$, and $\zeta_{nS}(\lambda)$ 
depend on $\lambda$, the decay rate is completely determined by $\lambda$.
Determination of the best value for $\lambda$ can be made by imposing
a requirement that the resummed formula for the decay rate should reproduce 
the measured rate. Therefore, the leading-order and the relative-order-$v^2$
NRQCD matrix elements are completely determined~\cite{Bodwin:2007fz}.

\section{Numerical Results \label{sec:result}}
 \begin{table}
 \caption{\label{table:param}
Scaled energy eigenvalues $\zeta_{nS}$ and $F_{nS}$ of the $S$-wave
heavy quarkonium with radial quantum numbers $n=1$, $2$, and $3$
as functions of the Coulomb strength parameter $\lambda$.}
 \begin{ruledtabular}
 \begin{tabular}{crrrrrr}
$\lambda$ & 
$\zeta_{1S}$\mbox{\hspace{2.7ex}}& $F_{1S}$\mbox{\hspace{1.2ex}}&
$\zeta_{2S}$\mbox{\hspace{2.7ex}}& $F_{2S}$\mbox{\hspace{1.2ex}}&
$\zeta_{3S}$\mbox{\hspace{2.7ex}}& $F_{3S}$\mbox{\hspace{1.2ex}}
\\
\hline
0.0&   2.338107&1.1248&4.087949&0.8237&5.520560&0.6983\\[-0.8ex]
0.1&   2.253678&1.1869&4.029425&0.8525&5.473169&0.7178\\[-0.8ex]
0.2&   2.167316&1.2532&3.970286&0.8821&5.425462&0.7375\\[-0.8ex]
0.3&   2.078949&1.3237&3.910531&0.9125&5.377441&0.7576\\[-0.8ex]
0.4&   1.988504&1.3989&3.850160&0.9437&5.329112&0.7780\\[-0.8ex]
0.5&   1.895904&1.4789&3.789174&0.9756&5.280478&0.7987\\[-0.8ex]
0.6&   1.801074&1.5641&3.727575&1.0083&5.231545&0.8196\\[-0.8ex]
0.7&   1.703935&1.6546&3.665364&1.0417&5.182316&0.8408\\[-0.8ex]
0.8&   1.604409&1.7507&3.602543&1.0758&5.132798&0.8622\\[-0.8ex]
0.9&   1.502415&1.8527&3.539116&1.1105&5.082996&0.8837\\[-0.8ex]
1.0&   1.397876&1.9608&3.475087&1.1459&5.032914&0.9054\\[-0.8ex]
1.1&   1.290709&2.0753&3.410458&1.1818&4.982560&0.9272\\[-0.8ex]
1.2&   1.180834&2.1965&3.345233&1.2183&4.931938&0.9492\\[-0.8ex]
1.3&   1.068171&2.3246&3.279418&1.2552&4.881053&0.9712\\[-0.8ex]
1.4&   0.952640&2.4599&3.213016&1.2927&4.829913&0.9933\\[-0.8ex]
1.5&   0.834162&2.6026&3.146031&1.3306&4.778522&1.0155\\[-0.8ex]
1.6&   0.712658&2.7529&3.078468&1.3689&4.726886&1.0377\\[-0.8ex]
1.7&   0.588049&2.9110&3.010330&1.4077&4.675010&1.0598\\[-0.8ex]
1.8&   0.460260&3.0773&2.941621&1.4468&4.622899&1.0820\\[-0.8ex]
1.9&   0.329215&3.2518&2.872344&1.4862&4.570560&1.1041\\[-0.8ex]
2.0&   0.194841&3.4348&2.802503&1.5260&4.517996&1.1262\\[-0.8ex]
2.1&   0.057065&3.6264&2.732099&1.5662&4.465212&1.1482\\[-0.8ex]
2.2&$-$0.084182&3.8269&2.661134&1.6066&4.412212&1.1702\\[-0.8ex]
2.3&$-$0.228969&4.0364&2.589611&1.6474&4.359001&1.1921\\[-0.8ex]
2.4&$-$0.377362&4.2550&2.517529&1.6885&4.305582&1.2139\\[-0.8ex]
2.5&$-$0.529425&4.4829&2.444888&1.7299&4.251959&1.2356\\[-0.8ex]
2.6&$-$0.685221&4.7202&2.371688&1.7716&4.198135&1.2573\\[-0.8ex]
2.7&$-$0.844808&4.9670&2.297928&1.8137&4.144112&1.2788\\[-0.8ex]
2.8&$-$1.008244&5.2235&2.223605&1.8561&4.089893&1.3002\\[-0.8ex]
2.9&$-$1.175584&5.4897&2.148718&1.8989&4.035481&1.3216\\[-0.8ex]
3.0&$-$1.346882&5.7657&2.073261&1.9421&3.980877&1.3428
 \end{tabular}
 \end{ruledtabular}
 \end{table}

\begin{figure}[t]
\begin{center}
\includegraphics[width=11.0cm]{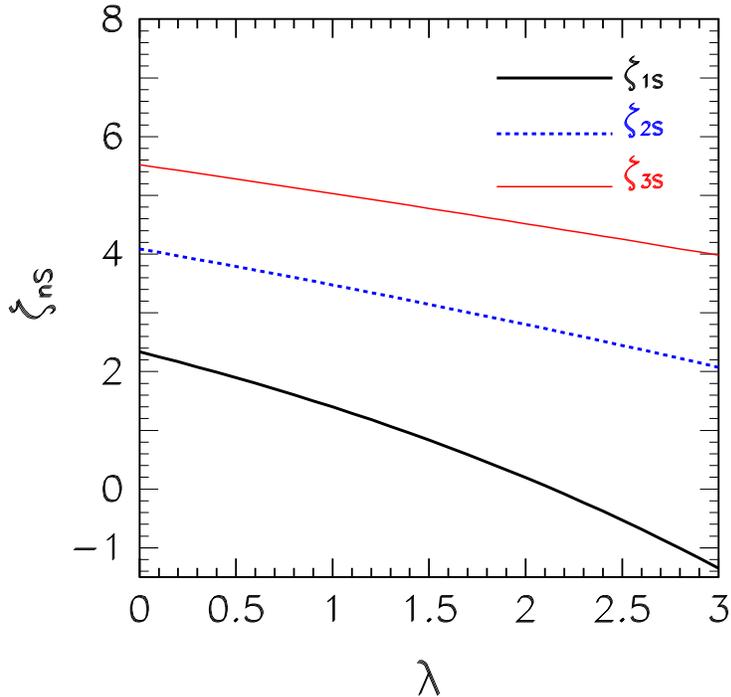}
\vspace{0.2cm}
\caption{Eigenvalues $\zeta_{nS}$ as functions of $\lambda$.
}
\label{figure1}
\end{center}
\end{figure}

\begin{figure}[t]
\begin{center}
\includegraphics[width=11.0cm]{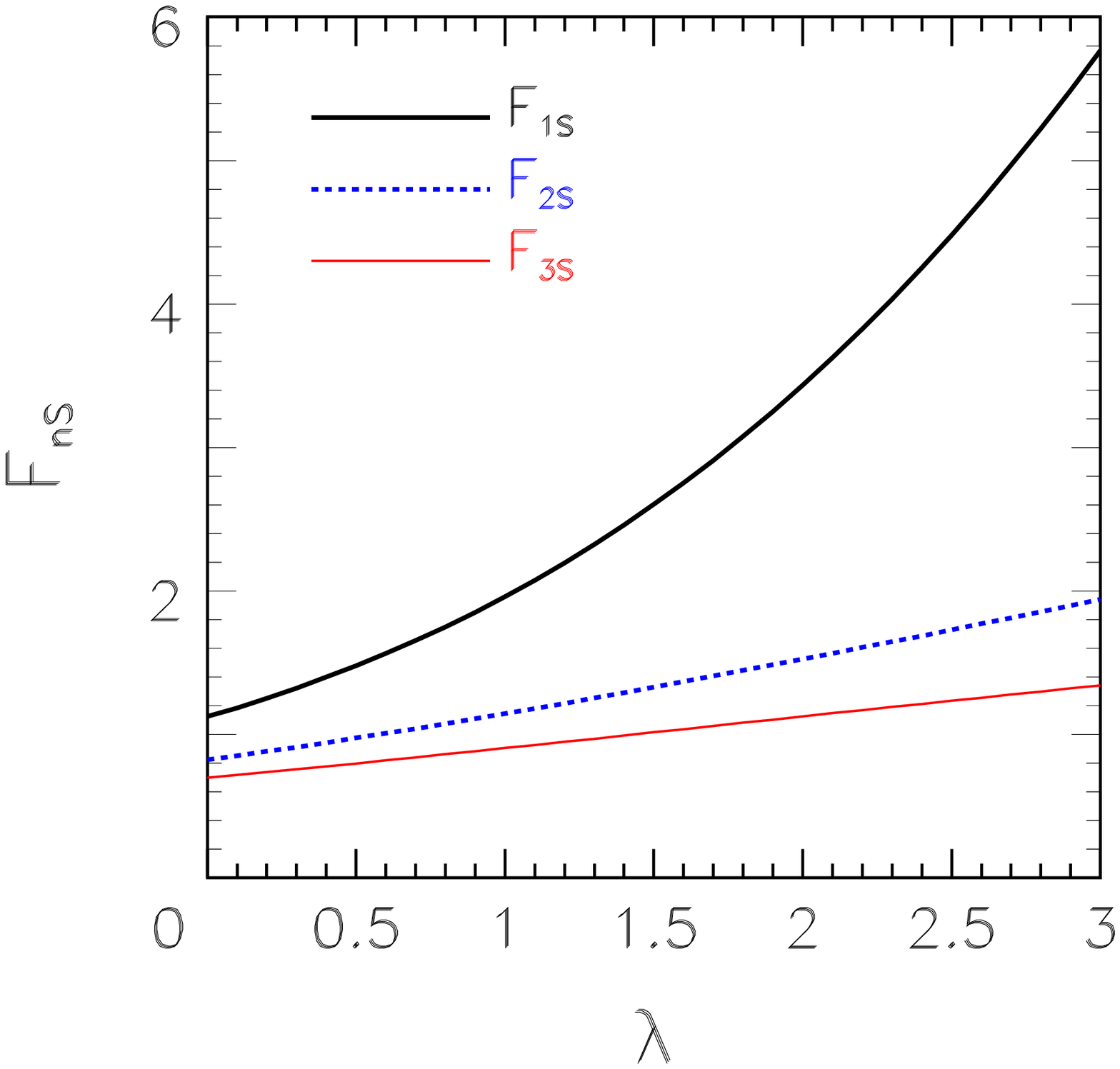}
\vspace{0.2cm}
\caption{$F_{nS}$ as functions of $\lambda$.
}
\label{figure2}
\end{center}
\end{figure}

In this section, we list our numerical values for the parameters and
derived quantities for the $S$-wave heavy quarkonia with radial
quantum numbers $n=1$, $2$, and $3$. In order to provide a set of
parameters that can be used for both charmonia and bottomonia,
we list the values for $\zeta_{nS}$ and $F_{nS}$ that are common
to both cases. In Table \ref{table:param}, we  tabulate those values
as functions of the model parameter $\lambda$, whose range
has been chosen so that the parameters can be used to determine
the NRQCD matrix elements for both charmonia and bottomonia.
The dependence of the eigenvalues $\zeta_{nS}$ and $F_{nS}$ on $\lambda$ 
are shown in Figs.~\ref{figure1} and \ref{figure2}, respectively.

In Ref.~\cite{Kang:2006jd}, the authors solved the Schr\"odinger equation
in Eq.~(\ref{radial}) numerically by using the inverse iteration 
method introduced in Ref.~\cite{Aichinger:2005}. 
The method uses a trial wave packet which is a 
linear combination of the eigenfunctions for a given system and 
the amplification operator of the form $(H-\zeta_{\rm trial})^{-1}$,
where $H$ is the Hamiltonian of the system and
$\zeta_{\rm trial}$ is a trial eigenvalue. If this operator acts on
an eigenfunction of the Hamiltonian, then the operator yields 
a multiplicative factor $(\zeta-\zeta_{\rm trial})^{-1}$,
where $\zeta$ is the eigenvalue of that eigenfunction.
As $\zeta_{\rm trial}$ gets closer to $\zeta$, the multiplicative
factor blows up. Therefore, if this operator acts on
a wave packet, then an eigenfunction, whose eigenvalue is the nearest 
to $\zeta_{\rm trial}$, is selectively amplified. Therefore,
by applying the operator repeatedly, one could obtain the eigenfunction
and the corresponding eigenvalue from the expectation value of the
Hamiltonian. By varying $\zeta_{\rm trial}$, one could also obtain
the energy spectrum. The details of the numerical method can be found in
Refs.~\cite{Kang:2006jd,Aichinger:2005}.
Here, we use this method to generate Table \ref{table:param}.

If one substitutes $\zeta_{nS}(\lambda)$ listed in Table \ref{table:param}
and the measured $1S$-$2S$ mass splitting for either the charmonium
or the bottomonium into Eq.~(\ref{m-lam}), one can obtain the mass
parameter $m(\lambda)$ as a function of $\lambda$.
$\epsilon_{nS}$ and  $\psi_{nS}(0)$ are calculable by substituting
the parameters $\zeta_{nS}(\lambda)$, $m(\lambda)$, and 
$F_{nS}(\lambda)$ into Eqs.~(\ref{eq:ezeta}) and (\ref{eq:psi0rhom2}), 
respectively. A convenient way to determine the optimal value for
the parameter $\lambda$ is given in 
Refs.~\cite{Bodwin:2006dn,Bodwin:2007fz}.

\section{Discussion \label{sec:discussion}}

We have computed derived quantities for various values of the model parameter
of the Cornell potential model for the $S$-wave heavy quarkonia
with radial quantum numbers $n=1$, $2$, and $3$. The scaled Schr\"odinger
equation is solved numerically for various values for the model parameter
$\lambda$ which determines the strength of the Coulomb potential relative
to the linear potential. The scaled energy eigenvalue $\zeta_{nS}$ and
a derived potential quantity $F_{nS}$ are computed as functions of
$\lambda$. These numbers are useful in determining the leading and 
relative-order-$v^2$ NRQCD matrix elements for both $S$-wave charmonia
and bottomonia with radial quantum numbers $n=1$, $2$, and $3$.
As an application, the leading-order NRQCD matrix elements for the
$\Upsilon(nS)$ have already been calculated and used to determine
the branching fractions and charm-momentum distributions for the
inclusive charm production in $\Upsilon(nS)$ decays, which are being
analyzed by the CLEO Collaboration~\cite{Kang:2007uv}. The result
listed in this paper can also be used to determine the NRQCD matrix
elements for the $2S$ charmonium states.  
Once recent studies~ \cite{Chung:2007ke,Lee:2007kf,Lee:2007kg}
to compute relativistic corrections to the leptonic width of the $S$-wave
spin-triplet quarkonium are extended to complete the resummation of
relativistic corrections to the process at next-to-leading order in 
$\alpha_s$, our results can also be used to determine the leading-order
and relative-order-$v^2$ NRQCD matrix elements for the $S$-wave quarkonium
with accuracies better than the best available values in 
Ref.~\cite{Bodwin:2007fz}.

\begin{acknowledgments}
We wish to thank Geoff Bodwin for checking the values of the parameters for
the $1S$ state, which were published in Refs.\cite{Bodwin:2006dn,%
Bodwin:2007fz}, and useful discussions.
We also express our gratitude to Eunil Won for a useful discussion on the
numerical method in Ref.~\cite{Kang:2006jd}. Critical reading of the
manuscript by Chaehyun Yu is also acknowledged.
The work of HSC was supported by the Korea Research Foundation
under MOEHRD Basic Research Promotion grant KRF-2006-311-C00020.
The work of DK was supported by the Basic Research Program of the
Korea Science and Engineering Foundation (KOSEF) under Grant No.
R01-2005-000-10089-0.
The work of JL was supported by the Korea Research Foundation
under Grant No. KRF-2004-015-C00092.
\end{acknowledgments}

{}


\end{document}